\pdfoutput=1
\RequirePackage{fix-cm}

\documentclass[smallextended]{svjour3}       
\smartqed  
\usepackage{graphicx}
\usepackage{cite}
\usepackage{amssymb,amsmath,amscd, mdwmath}

\usepackage{amsfonts}
\usepackage{amsmath}
\begin{document}

\title{
Feature analysis   of    multidisciplinary scientific  collaboration patterns   based on  PNAS
}


\author{ Zheng Xie\and  Miao Li\and Jianping Li \and Xiaojun Duan\and Zhenzheng Ouyang 
}


\institute{
  Z. Xie$^*$ Correspondence: \email{xiezheng81@nudt.edu.cn}, J. Li \email{jianpingli65@nudt.edu.cn},   X. Duan \email{xjduan@nudt.edu.cn}, Z. Ouyang \email{zzouyang@nudt.edu.cn}\at
               College of Science, National University of Defense Technology, Changsha, 410073,  China  \\
                             \\  M. Li \email{limiaojoy@sjtu.edu.cn} \at   School of Foreign Languages, Shanghai Jiao Tong University, Shanghai, 200240, China \\
                           The first three authors have  contributed equally to this work.
\\
 }

\date{Received: date / Accepted: date}

\maketitle

\begin{abstract}
The features of collaboration patterns  are often considered to be different from discipline to discipline. Meanwhile, collaborating among disciplines is an obvious feature emerged in modern scientific research, which incubates several interdisciplines. The features of collaborations in and among the disciplines of biological, physical and social sciences are analyzed based on 52,803 papers published in a multidisciplinary journal PNAS during 1999 to 2013. From those data, we found similar transitivity and assortativity of collaboration patterns  as well as the identical distribution type of collaborators per author and that of papers per author, namely a mixture of generalized Poisson and power-law distributions. In addition, we found that interdisciplinary research is undertaken by a considerable fraction of authors, not just those with many collaborators or those with many papers.
 This case study provides a window for understanding aspects of multidisciplinary and interdisciplinary collaboration patterns.

\end{abstract}

\section{Introduction}


Natural  and social sciences provide    methodical approaches to study, predict and explain
   natural
phenomena and sociality (human behaviors and psychological states) respectively\cite{Weingart}.  The specialization of knowledge    in these sciences forms
various disciplines. Meanwhile,
 to   solve problems whose solutions are beyond  the scope of a
single discipline,
researchers     need  to integrate
data, techniques, concepts, and theories from several disciplines\cite{Cooper,  Hurd,  National,  Hirsch}.   Interactions between disciplines incubate    several interdisciplines,   fuzz  the boundary of  natural and social  sciences, and produce
many important  scientific   breakthroughs\cite{Liu,Siedlok,Gooch}.

Studying   collaboration patterns      within and across disciplines or sciences
  contributes to understand    the diversity of   cooperative behaviors    and fusion modes of  knowledge.
  Papers  of  multidisciplinary journals provide  an informative and reliable platform for this studying,
   because the media of natural and social  sciences  mainly count on papers\cite{Lariviere2,Moody, Glanzel,Hicks}.
 Here we  investigated
    the    patterns
based on   52,803
papers    published  in  {\it  Proceedings of the National Academy of Sciences}~(PNAS) over the years 1999-2013.$^a$
 The content of dataset spans  three science categories: social sciences and two principal  sub-sciences in natural sciences, viz. biological   and physical sciences.


     Collaboration relationship can be expressed by graphs, termed as coauthorship networks.
     Hence the patterns can be studied in   network perspective.
Coauthorship networks from different scientific fields appear specific similarities, such as partial transitivity of coauthorship, homophily on the number of collaborators, the right-skewed distribution of collaborators  per author\cite{Newman1,Newman4,Barab,Newman0,Sarigol,Xie3,Tomasello}.
    These commonalities  also appear in
the collaboration networks of   three author sets  (which come from the three science categories  of PNAS  respectively).
We dived more into the rule and reason   of  these  commonalities.
 We found  that  the   distribution of collaborators   per author and that of papers per author
   follow the same distribution type: a mixture of a generalized Poisson distribution and a power-law.
   We provided a possible explanation  for
the distribution type and   these  commonalities through
   the   diversity of   author abilities to  attract   collaborations.



A range of previous works discussed    quantitative indexes of
interdisciplinarity for sciences\cite{Braun,Porter1,Levitt2}, for disciplines\cite{Porter,Chen,Rafols,Abramo}, for universities\cite{Bordons},  for journals\cite{Leydesdorff,Zhang}, and for research teams\cite{Lungeanu}. Some works   addressed
the correlation between
 interdisciplinarity and scientific impact\cite{Lariviere,Lariviere1,Rinia,Wang} (e.~g.   citation catching ability\cite{Levitt1,Levitt,Chen1}).
Based on  specific general ideas of these references, we studied
   interdisciplinary  activities of  PNAS through  paper co-occurrence of  disciplines, and through some indexes
  calculated  based on the co-occurrence, such as
      Rao-Sterling diversity\cite{Stirling}, and   betweenness centrality\cite{Leydesdorff2007}.

 We further studied the collaboration patterns    across disciplines, and
    found that  a considerable  proportion of authors and papers in physical and social sciences involved in interdisciplinary research.
  The multidisciplinary coauthorship network extracted from the data  has  a giant component, which contains more than 88\%, 80\% and 71\% authors in biological, physical and social sciences   respectively.
 A considerable  number of authors contribute to the formation of giant component. The
 contributions of
   author activity and productivity to the formation  increase  over time.
The high extent of interdisciplinarity shown by the case study might not be  representative of general collaboration patterns, because
authors could   submit more interdisciplinary work to  multidisciplinary journals than domain specific ones.



This report is structured as follows: the data processing is described in Section 2; the similarities and interactions are analyzed in Section 3; and the discussion and conclusion are drawn in Section 4.

\section{The Data  }

\subsection{Reason for using the data  }

The case study   involves  two concepts, namely
multidisciplinarity (researchers from different disciplines study
  within their
disciplines) and interdisciplinarity  (study beyond disciplinary  boundaries)\cite{Besselaar}.
Multidisciplinarity could be viewed as a combination of disciplines, and
interdisciplinarity as a merging of them.
 A multidisciplinary journal  with the scope covering  natural and social sciences can be utilized to
 analyze
   the interactions between    science categories.
Such journal can be also utilized to compare
the  collaboration patterns of  multi-disciplines  and
 find   similarities.
PNAS   publishes high  quality  research papers, and provides   reliable discipline  information of those papers.
 The journal also provides  a high quality data platform  for analyzing worldwide collaboration patterns, because  nearly half of   its papers come from authors outside
the United States.

Multidiscipline journals: {\it Science},	{\it Nature}   and {\it Nature Communications} do not provide discipline information of papers.
{\it Journal of the Royal Society Interface}    focuses  on the cross-disciplinary research at the interface between the physical and life sciences, but does not involve social sciences.
Our analysis  is restricted to PNAS, which brings    limitations to our findings.
For example,        the   media of social sciences   not only  count  on    papers,  but also  on books\cite{Glanzel,Hicks}.
Hence the results obtained
must be carefully interpreted as being the patterns of researchers who publish papers in the chosen journal.
However,  due to the influence and representability of PNAS,   the case study could
contribute  to understanding    aspects of multidisciplinary and interdisciplinary collaboration patterns.

\subsection{Discipline  information}

Most   papers of the dataset have been      classified into three first-class disciplines (biological, physical, and social sciences) and $39$ second-class disciplines (Table~\ref{table1}).
  Interdisciplinary papers are classified into several   disciplines.
 The data contain  43,304  biological papers (including 3,957 papers of biophysics), which account for   82.01\% of the total.
 The data  also contain  5,987 physical papers  and 1,310  social papers.    There are 2,961    interdisciplinary papers belonging to more than  one of the second-class disciplines, which account  for $5.61\%$ of the total. The significant difference  of discipline proportion  does not mean the preference  for PNAS. In reality, the number of researchers  involved in  natural sciences (especially,   biological sciences) is far more than that  of researchers involved in social sciences\cite{Kagan}.
  There are 1,842  papers that
are only classified into the first-class disciplines. For these papers,    their second-class discipline are   regarded to be missing,
but which have been  regarded to be    the same as their
first-class disciplines in our pervious work\cite{Xie4}.
Hence the data in Table~\ref{table1} are different from  those in   Reference \cite{Xie4}.

\begin{figure}
\includegraphics[height=3.1   in,width=4.7    in,angle=0]{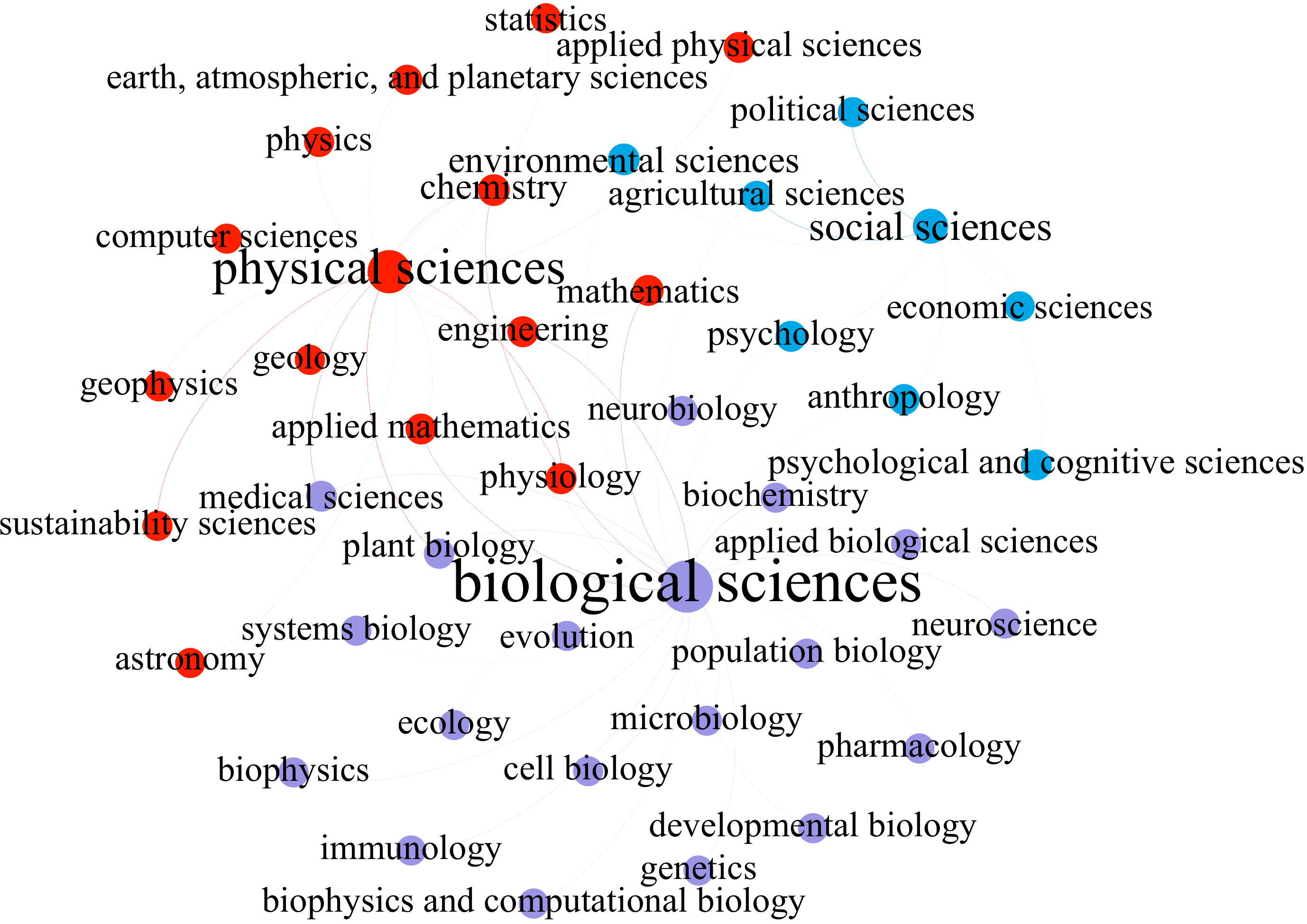}
 \caption{  {\bf The    relationship  between the first-class  and the second-class disciplines.  } The network is built based on the discipline information of papers in PNAS  1999-2013.  Two disciplines are connected if they are the first-class and the second-class disciplines of  a paper.
 The node size   indicates node degree.
 }
 \label{fig2}      
\end{figure}

\begin{figure}
\includegraphics[height=2.6   in,width=4.4     in,angle=0]{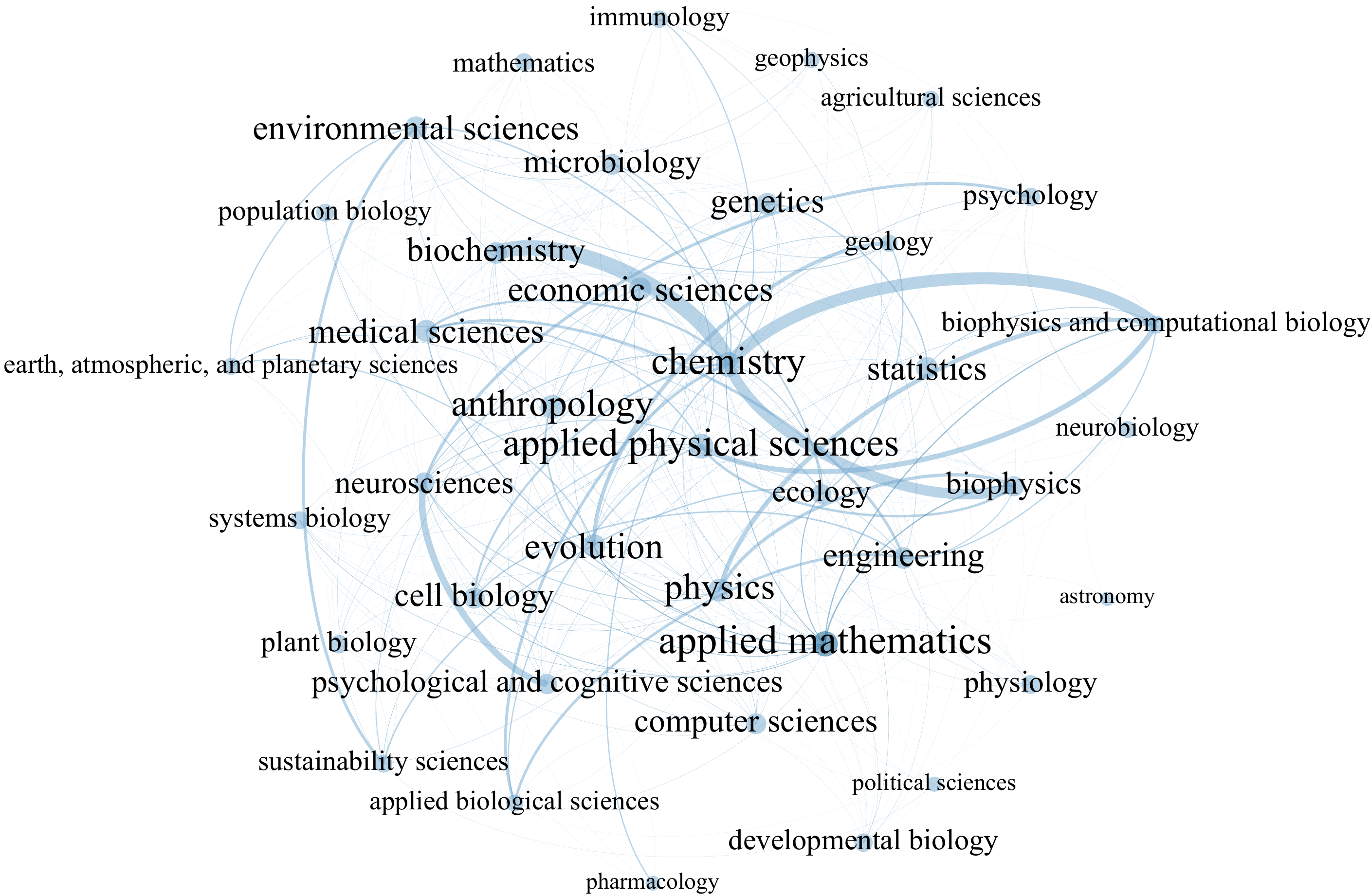}
 \caption{  {\bf Interactions between the second-class disciplines.  } The weighted network is built based on the discipline information of interdisciplinary papers in PNAS 1999-2013.   Edge width indicates edge weight: the number of interdisciplinary papers between two connected disciplines.  
 }
 \label{fig22}      
\end{figure}

Based on the discipline information of   papers, we   constructed a network to express  the relationship
between the first-class and the  second-class disciplines~(Fig.~\ref{fig2}),   where  two disciplines are connected if they are the first-class and the second-class disciplines of  a paper.
We  can also construct  a network to express  the interactions between the second-class disciplines (Fig.~\ref{fig22}),
  where each node is a  discipline
 and two nodes are connected if there is a paper belonging to them simultaneously.
 These networks could evolve with the discipline information of newly published papers. So using the latest data, one may have a more comprehensive view.

\begin{table}[ht] \centering \caption{ {\bf  Specific  indexes of the second-class   disciplines  in PNAS 1999--2013.}
}{\footnotesize
\begin{tabular}{  l rrrrrr} \hline
Disciplinary &$m$ &$n$     &  $k_1$   & $k_2$ & $b$ \\ \hline
Agricultural science &22 &226& 9&20& 3.19 \\
Anthropology &114& 556&24& 110 & 40.02  \\
Applied biological science& 135& 767&9&134 &1.79\\
Applied mathematics &191& 380&27&182&49.39\\
Applied physical science &309& 816&26&299&29.14\\
Astronomy &3& 50&3&3&0.13\\
Biochemistry& 333 &6,303&19&327&16.96\\
Biophysics& 359& 3,957&16&359&7.91\\
Biophysics and computational biology &468& 1,532& 11&467&7.95\\
Cell biology& 135& 3,717&18&130&12.71\\
Chemistry& 1,003& 8,645&26&1,003&49.73\\
Computer science& 77& 101&  17& 70& 9.50\\
Developmental biology &33& 1,525&12 &30&1.66\\
Earth, atmospheric, and planetary sciences &78& 243& 9&77&1.58\\
Ecology& 162& 1,084&15&162&10.00\\
Economic science& 94& 171&21&94&20.88\\
Engineering &217& 392&19&217&13.85 \\
Environmental science &184& 695&20&183&25.44\\
Evolution& 233& 2,274&22&216&25.81\\
Genetics &103& 2,664&20&97&12.68\\
Geology &137& 285&10&136&2.79\\
Geophysics& 23& 175&7&23&1.51\\
Immunology& 43& 3,070&10&38&1.45\\
Mathematics& 18 &561&11&17&3.36\\
Medical science& 181& 4784&20&170&14.01\\
Microbiology &92& 2,812&17&89&11.85\\
Neurobiology &16& 1,003&9&16&0.87\\
Neuroscience& 290& 4,398&16&280&12.00\\
Pharmacology &26 &594&4&26&0.08\\
Physics &229& 4,818&22&227&18.24\\
Physiology &33& 1,116&12&32&5.82\\
Plant biology& 27 &1,700&12&27&4.62\\
Political science& 7& 17&5&7&0.54\\
Population biology &27 &166&11&26&4.04\\
Psychological and cognitive science &160& 487&16&159&5.09\\
Psychology &83 &449&12&83&3.62\\
Statistics &90 &146&20&85&19.34\\
Sustainability science& 123 &399&11&120&7.66\\
Systems biology &36 &159 &11&36&1.80\\
\hline
 \end{tabular}}
 \begin{flushleft}
    The number of  papers $n$ and    that of  interdisciplinary  papers $m$ of a discipline
   are counted based on the discipline information provided by PNAS.
       The   degree $k_1$, weighted degree $k_2$,   and betweenness centrality $b$ of a  discipline are calculated based on  the  weighted  network in Fig.~\ref{fig22}.
 \end{flushleft}
 \label{table1}
\end{table}

\subsection{Coauthorship}

Identifying   ground-truth authors, termed as disambiguating author names,
  is an important, time-consuming, but a necessary procedure of coauthorship analysis.
Several  methods   use the information of the provided
names on papers (e.~g. initial based methods\cite{Milojevic2}).
The dominant  misidentification
of initial based methods is caused by
    merging two or more different authors as one.
 Hence, it
deflates the number of unique authors,  and inflates   the size of the ground-truth giant component.
 Requiring additional information (e.~g. email address) helps to  reduce merging errors, but brings   the difficulty
of collecting   information.

In   PNAS 1999-2013,  $93.1\%$ authors provide full   first name. So
 the provided names on papers are directly used to identify authors.
However,
utilizing  surname and the initial of the first given
name  will generate  a lot of merging errors of  name disambiguation\cite{Kim1}.
The proportion
of these  authors
 in the data is $2.9\%$, and the proportion of these authors further  conditioned on publishing  more than one paper is $0.3\%$.
  Meanwhile,   even  utilizing   full   names
     still produces merging  errors, if   some authors provide exactly the same name.
  Chinese names were found to account for name repetition\cite{Kim1}.
We calculated the proportion  of the names with a given name less than six characters and a surname among major 100    Chinese surnames.$^b$
   The      proportion  of these authors in the data  is $2.7\%$,
    and that of these authors further  conditioned on publishing  more than one paper is $1.1\%$.
 The small values of these four proportions
show  that
the impact  of
name repetition is
limited.
 These proportions for specific  subsets of the data are listed in Table~\ref{tab21}.

\begin{table*}[!ht] \centering \caption{{\bf Specific   statistical   indexes of  the analyzed networks.} }
\begin{tabular}{l r r r r r r r r r rr} \hline
Data&     $a$& $b$  & $c$    &$d$    \\ \hline
  PNAS 1999-2013    &   2.9\%&   0.3\%   & 2.7\%&1.1\%\\
Biological sciences   &  2.7\% & 0.2\%    & 2.7\%&1.1\%  \\
Physical sciences     & 4.8\%&0.4\%  &  4.4\%&0.9\% \\
Social sciences  & 2.3\%&  0.1\%     &  2.2\% &0.3\%\\
Biophysics   &  4.1\%&0.3\% &4.0\%&1.0\%\\
Interdiscipline  &2.6\%&0.1\%  &3.6\%& 0.6\%\\
\hline
 \end{tabular}
  \begin{flushleft} Indexes $a$ and $b$ are
    the proportion  of the  authors only providing the initial of their first given name  and their surname, and that of these authors further conditioned on publishing  more than one paper  respectively.
  Indexes $c$ and $d$ are     the proportion of the authors with a surname  among the major 100    Chinese surnames and a given name less than six characters, and that of these authors further conditioned on publishing  more than one paper  respectively.
\end{flushleft}
\label{tab21}
\end{table*}

The  method   adopted here   will split one author as two or more, if the author does not provide his   name  consistently.
       Splitting       underestimates
          the
giant component size,
     and  the indexes used as evidences for   universality of interdisciplinary research.
Hence the results   in  Subsection  3.5, 3.6 could be regarded as conservative ones.
In addition,  the inaccuracy caused by the adopted  method   does not change
  the  ground truth distribution type of  collaborators per author and  that of papers per author\cite{Kim1}.



\section{Data analysis }

\subsection{Network properties}

Coauthorship is a $n$-ary relation, $n\in \mathbb{Z}^+$, hence it  can be   expressed by a  hypergraph, a generalization of a graph in which an edge (termed as hyperedge) can join any number of nodes.
  Represent authors as nodes, and  the  author group of  each paper (paper team) as   a hyperedge.
Then we can extract
a coauthorship network   from  a hypergraph as a simple graph,  where   edges are formed between every two nodes in each hyperedge, and the multiple edges are treated as one. The   terms ``degree" and ``hyperdegree" for nodes  are used to express
the number of collaborators   and that of  papers for   authors   respectively.

   The data show that the average paper team size   of biological sciences (6.624) and that of physical sciences (5.254) are larger than that of  social sciences (4.634). The size relation fits the reality  that  the sizes of research teams  are usually larger in natural sciences,   and smaller    in social sciences\cite{Kagan}.
Now let us  consider  the coauthorship networks of the considered papers in specific disciplines or science categories.
 All of these networks are   highly  clustered, assortative, and their average shortest
path length   scale as the logarithms of
their number of nodes~($\log$NN$\approx$AP in Table~\ref{tab2}). These properties do not mean all of the networks are small-world. The network  of social sciences is an exception, which even has no   component containing more than 10\% authors. However, it does not mean that the research in social sciences is carried out in isolation. In fact,
 71.5\% authors in social sciences belong to the giant component of the coauthorship network generated by the whole data.
Therefore,  analyzing the collaborations of authors restricting in single discipline has limitations. So we proceeded the   analysis   in the environment of  all disciplines.
\begin{table*}[!ht] \centering \caption{{\bf Specific   statistical   indexes of  the analyzed networks.} }
\begin{tabular}{l r r r r r r r r r} \hline
Network&NN&NE   & GCC & AC &AP   & PG      \\ \hline
  PNAS 1999-2013  &202,664&1,225,176 &0.881 &0.230 & 6.422   &  0.868   \\
Biological sciences &  184,872&1,150,362 &0.881 &0.232&6.364 &  0.880\\
Physical sciences     &24,766 &101,166 &0.933&0.452 &10.89    & 0.455 \\
Social sciences & 5,121&18,786& 0.946 & 0.683 &6.574  &0.087 \\
Biophysics &13,480& 48,012&0.905&0.177&  7.665  &0.636 \\
Interdiscipline & 13,680& 53,588& 0.951 &0.558 &9.397    &0.093\\
\hline
 \end{tabular}
  \begin{flushleft} The indexes are    the number  of nodes (NN), the number  edges (NE),  global  clustering coefficient (GCC),  degree assortativity coefficient~(AC),   average shortest path length~(AP),  the node proportion of the giant component~(PG).
  The   AP of   the first two networks are approximately  calculated by sampling 400,000 pairs of nodes.
\end{flushleft}
\label{tab2}
\end{table*}

\subsection{Degree and  hyperdegree}
\label{sec3}



\begin{figure}
\includegraphics[height=3.    in,width=4.7    in,angle=0]{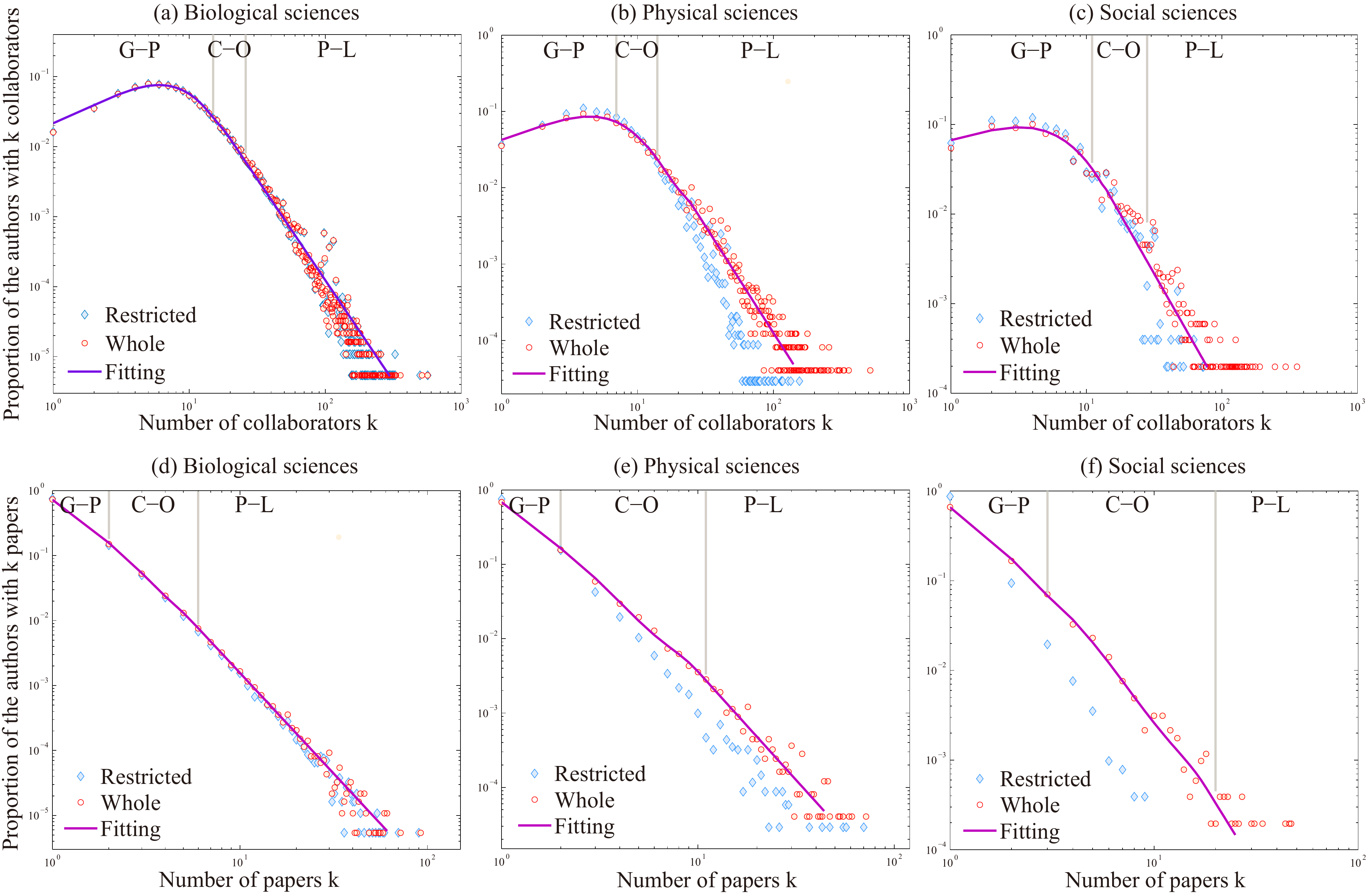}
 \caption{ {\bf Distributions of collaborators/papers per  author.}
The panels   show the distributions counted in PNAS 1999--2013 (red plots), and   those  counted in the papers of each    science category (blue diamonds).     Fitting distributions (purple curves) are    mixtures   of a generalized Poisson distribution and a power-law distribution. Fitting parameters  are listed in Table~\ref{tab3}.
The regions ``G-P",  ``C-O", ``P-L" stand  for generalized Poisson, cross-over  and power-law respectively.
 }
 \label{fig4}      
\end{figure}

Aggregate degree and hyperdegree on   the data (not restricted  in single science category),
and observe
  the degree distributions    and hyperdegree distributions
 of three author sets (which come from the three science categories respectively).
We   found that although  collaboration level differs from one science category  to another,
all of the  distributions  emerge  a hook head, a fat tail, and a cross-over  between them, which could be viewed as a common feature of coauthorship networks~(Fig.~\ref{fig4}).   The head and tail can be   fitted by log-normal distribution and power-law distribution respectively\cite{Milojevic1}.

 These distributions  can also  be  fitted, as a whole, by a mixture
of a generalized Poisson distribution and a power-law distribution.
 The fitting parameters   are listed in  Table~\ref{tab3}.
 We
performed  a two-sample Kolmogorov-Smirnov (KS) test to compare the distributions of    two data vectors:
node indexes (i.~e. degrees, hyperdegrees), the
samples drawn from the corresponding fitting distribution.
  The null hypothesis is that the two data vectors are from the same distribution.
 The $p$-value of each fitting shows
 the test cannot reject  the null hypothesis at   5\% significance level.
Note that  $\chi^2$ goodness-of-fit test is not suitable here, due to the small number of large degree authors.



Regarding authors  as samples,
  a mixture distribution means those samples come from different populations, namely the collaboration patterns  of the authors with few collaborators and papers differ  from
those with many collaborators and papers. In Reference\cite{Xie6},
a  possible  explanation (which is free of disciplines)  is given  for   the emerged  mixture type  of empirical    degree  distributions.
With the same
general ideas, a similar  explanation   can be adopted    for   hyperdegree distributions  as follows.

 The event whether a researcher collaborates  with one another  to publish a paper   can be regarded as  a  ``yes/no"  decision.
  So the hyperdegree of a researcher  is equal to the number of successes in a sequence of     decisions made by the candidates who want  to coauthor with that researcher.
      Suppose the   number of those candidates  to be $n$.  Suppose the collaboration  probability of each candidate  to be  $p$.
  Then,   the hyperdegrees will  follow  a binomial distribution  $B(n, {p})$, and so a Poisson distribution  with  expected value   $n {p}$ approximatively~(Poisson limit theorem).
 The   value  of  $n {p}$ varies from   author to author, due to the diversity of authors' ability to attract collaborators.

Decisions of authors could be dependent. For example, collaborating with the researchers who have publishing experience helps to publish a paper. Hence  we could regard hyperdegree as a random variable following a generalized Poisson distribution (which allows the occurrence probability of an event to involve memory\cite{Consul}). In empirical data, most hyperdegrees are around their mode. Hence  we could think of that they follow some generalized Poisson distributions with an expected value around their mode, and so form the generalized Poisson part of a hyperdegree distribution. A few authors experience a cumulative process of papers, which makes a hyperdegree distribution skew to the right and form a fat tail.







\begin{table*}[!ht] \centering \caption{{\bf  The parameters  of fitting functions.} }
\begin{tabular}{l r r r rr r r r r r r r} \hline
Degree distribution &$a$ &$ b$   & $c$ &$d$  &$s$ & $B$   & $E$   & $G$ &$P$  &$p$-value  \\ \hline
Biological sciences  &4.843 &0.464 & 74.27 & 2.889 &  1.049  & 15& 26 &20& 50&         0.203
\\
Physical sciences&  3.958&  0.477  & 49.31   &2.798 &1.037 &7 &14    &20&53& 0.178 \\
Social sciences & 3.292& 0.513  & 20.78 &2.657  &1.046   &  11 &28 &20&35 &    0.111
  \\\hline
Hyperdegree distribution &$a$ &$ b$   & $c$ &$d$  &$s$ & $B$   & $E$   & $G$ &$P$  &$p$-value  \\ \hline
Biological  sciences &0.028 & 0.269 & 1.968&3.099 &35.57 & 2 &6 &10&13& 0.979  \\
Physical  sciences &0.021  &  0.320 & 2.977 &2.916& 47.15& 2 &11&10 &10& 0.625  \\
Social  sciences &  0.022& 0.375   & 19.48  &3.665& 46.24 & 3 &20 &10&11&    0.206\\
\hline
 \end{tabular}
  \begin{flushleft}
  The ranges of  generalized Poisson $f_1(x)$,  cross-over, and power-law $f_2(x)$  are $[1,E]$,  $[B, E]$, and  $[B,\max(x)]$ respectively.
  The fitting function is $f(x)=q(x)      s f_1(x) +(1-q(x))f_2(x)$, where
  $ q(x)=\mathrm{e}^{ - (x -B)/(E-x ) }$.    The fitting processes are:
  obverse proper $G$ and $P$; calculate parameters of $sf_1(x)$ (i.e. $a$,  $b$,    $s$) and $f_2(x)$ (i.e. $c$,    $d$)   through regressing  the empirical  distribution   in  $[1,G]$ and $[P,\max(x)]$ respectively;      find  $B$ and $E$ through exhaustion    to  make $f(x)$ pass  KS test ($p$-value$>0.05$).
 The sum   of each $f(x)$ over  $[1,\max(x)]$  is  near unity, which means that $f(x)$ can be regarded as a probability density function.
   \end{flushleft}
\label{tab3}
\end{table*}

\subsection{Transitivity   of    coauthorship}
\label{sec4}


Transitivity in society is that
``the friend of my friend is also my friend", which is a typical feature of social affiliation networks. In academic society, collaborators of an author likely acquaint and so coauthor with each other. For example, organizational and institutional contexts drive the formation of transitive coauthorship, and so contribute to
   clustering structures
  emerging  in coauthorship networks.





 The   transitivity of a    network can  be quantified   by two indexes in    graph theory, namely
 global  clustering
coefficient  (the fraction of connected triples of nodes
which also form ``triangles") and local clustering coefficient  (the probability  of a node's two neighbors  connecting).
High transitivity is   a common feature of  coauthorship  networks\cite{Newman1}.


 To what extent the transitivity  is due to the activity of
authors   in academic society?
 The activity  can be partly reflected through    the number  of  collaborators, namely degree.
 Hence,  the extent
can be sketched through the correlation coefficients between degree and local  clustering coefficient.
  Note that  the correlation coefficients
indicate the extent of a linear
relationship between two variables or   their  ranks. The   coefficients of variables $X$ and $Y$  generally do  not completely
characterize   correlation, unless   the conditional expected value of $Y$
given $X$, denoted by $E(Y|X)$, is   linear or approximate linear
function in $X$.
   The   conditional expected value   of local  clustering coefficient given degree
 is the average local clustering coefficient     of $k$-degree nodes, denoted    by    CC$(k)$.
  The    approximatively linear trend  of CC$(k)$~shown in Fig.~\ref{fig3}
  guarantees  the effectiveness of correlation analysis in Table~\ref{tab4}.
The decreasing trend   cannot be deduced out   from     degree information. The denominator of the local  clustering coefficient of a node grows quadratically with its degree, but the   numerator cannot be calculated   from   degree information.


\begin{table*}[!ht] \centering \caption{{\bf Correlation coefficients between  degree  and transitivity/clustering   indexes.} }
\begin{tabular}{l r r r r r r r r r} \hline
Discipline &Indicator  &Mean& Std&  SCC  &PCC      \\ \hline
  &LCC  &  0.860 & & -0.398 & -0.401 \\
Biological sciences   &LTC  &0.001 & 0.005 & 0.275&0.077\\
   & DN   & 21.09  & & 0.543 &0.400 \\
  & HN   & 3.015&15.47  &0.070& -0.046\\
\hline
  &LCC    &  0.806 & &-0.336 &  -0.382 \\
Physical  sciences &   LTC  &0.001 &0.005& 0.306&0.074\\
      &DN   & 15.48& &  0.625& 0.346\\
    & HN  &  2.682&12.44&0.169 &  0.015\\
\hline
   & LCC  & 0.784 && -0.177 &  -0.263 \\
 Social sciences &  LTC  &0.001 &  0.006&   0.292& 0.050\\
     &  DN  &  12.87&&   0.723& 0.482\\
    & HN  &  2.268&   10.89&0.175 &   0.030\\
\hline
 \end{tabular}
  \begin{flushleft}
   The indexes  are  local clustering coefficient~(LCC),  the local transitivity of collaboration~(LTC), the average degree  of node  neighbors~(DN), the average hyperdegree of node  neighbors~(HN).
We calculated  the mean  of these indexes over authors,
the Spearman  rank correlation coefficient~(SCC)  and   Pearson product-moment correlation
coefficient~(PCC) between each index and degree. For the two indexes with small PCC, we calculated their standard deviation (Std).
\end{flushleft}
\label{tab4}
\end{table*}






Does  the  decreasing trend  of CC$(k)$   mean activity depresses transitivity?  A positive answer to it is against common sense.
In PNAS 1999-2013, 74.62\% authors
only publish  one paper in the data, and
the paper team sizes of
99.9\%   papers follow a generalized Poisson part, namely are around the average paper team size  6.028.
 The boundary of generalized Poisson part  is detected by the boundary point  detection algorithm for  probability density functions  in Reference\cite{Xie6}  (listed in Appendix).
 Hence the local clustering coefficients of most  small degree authors are close to 1~(Fig.~\ref{fig3}).
A few    authors experience   a
long  period of collaborations, whose degree is obtained by accumulated over papers.
For these authors,   their  collaborators
in different papers could not   collaborate, which decreases their local clustering coefficient.
  Hence the puzzling thing does  not  contradict  with common sense, but   is due to insufficiency of measuring transitivity such a  dynamical property by   counting ``triangles" on a static network.

To design a more reasonable index measuring transitivity, let us come back to the original meaning of transitivity on coauthorship: the probability of two  collaborators
 (who do not coauthor yet) of a researcher coauthoring in   future. The probability can be calculated for dynamic hypergraphs of collaborations  through  time information.
Averaging   the probability over authors    measures the global  transitivity, the value of which  is quite low in each   science category (Table~\ref{tab4}).
Note that the calculation is limited  in PNAS 1999-2013, and transitivity may happen in other journals or in other time period.
So  the values of  transitivity   here may be underestimated.
 The    increasing trend   of  the transitivity probability    of $k$-degree authors    (TC$(k)$ in Fig.~\ref{fig3})   means the activity contributes to  transitivity.  It fits  common sense: a researcher  with many   collaborators  is   likely to introduce his collaborators  to cooperate.




%
%
%


%


\subsection{Homophily of coauthorship}


Coauthorship is based on specific    features of researchers in common, including interest and geography. The homophily phenomenon   appears in many social relations, and  is called   assortative mixing in network science~\cite{Newman4}.
Do  authors  of each  science category
  prefer  to coauthor with  others that are similar in
 social activity or productivity?
 The   social activity and productivity of authors
can be quantified    by two     indexes, namely degree and  hyperdegree respectively.
  Then the preference  of an index
 could be sketched through  the correlation coefficient between two variables, namely the  index  of a author  and the average index  of the author's neighbors.  Positive correlation means assortative, negative     disassortative, and zero no preference.

Degree assortativity is a  feature of coauthorship networks~\cite{Newman4}.
Does  it  mean  sociable researchers (with
many collaborators)  will  preferentially coauthor with   other sociable researchers, and   unsociable  to unsociable?
In a previous study~\cite{Xie7}, we   showed
that the proportion of top  5.99\% most sociable authors (measured according to degree) having coauthored with another such author is 99.5\%.
The proportion  may even be underestimated, because these authors probably coauthored before 1999 or in other situations.
Note that the splitting and merging errors of the used  name disambiguation method   affect the proportion at certain levels.
Even so, the proportion is still remarkable.

\begin{figure}
\includegraphics[height=3.    in,width=4.7   in,angle=0]{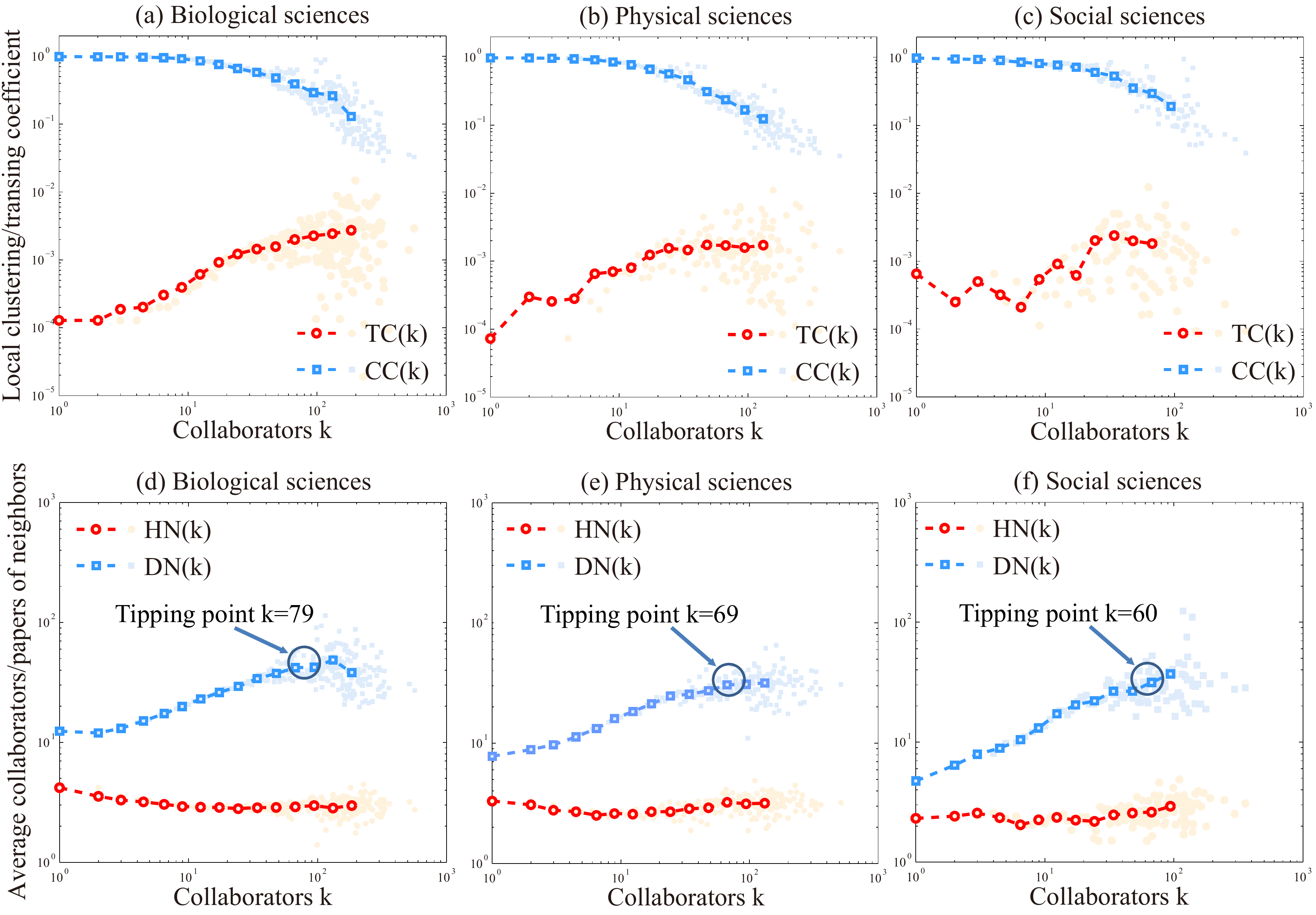}
 \caption{{\bf
 Conditional expected values
of specific indexes given degree.} From $k=1$ to $\max(\mathrm{degree})$, we average  over $k$-degree nodes   for local clustering coefficient~(CC$(k)$), the local  transitivity of collaborations~(TC$(k)$), the average degree of node   neighbors~(DN$(k)$), and the  average hyperdegree  of node neighbors~(HN$(k)$).  The data are binned on abscissa axes to
extract the trends hiding in noise. }
 \label{fig3}      
\end{figure}

However, if sociable researchers  only   coauthor with   sociable ones, then there will exist many  sociable researchers, which is against   empirical degree distributions.
Now let us  analyze the influence of the  social activity of authors   on  degree assortativity.
For the authors with  $k$-degree,  denote the average degree  of their neighbors by DN$(k)$.
 There exists a  trend change   in   DN$(k)$  of each empirical dataset: the head part has a clear increasing trend, but the tail part does not~(Fig.~\ref{fig3}).
It means that degree assortativity are mainly contributed by   small degree authors.

The tipping point  of the trend  of DN$(k)$
is detected by the boundary point  detection algorithm for general functions in Reference\cite{Xie6}  (listed in Appendix).
Inputs of the algorithm are DN$(k)$, $g(\cdot)=\log(\cdot)$  and $h(x)= a_1 x^3 + a_2 x^2 + a_3 x  + a_4$ ($x$, $a_i\in  {\mathbb{R}}$, $i=1,...,4$).
  Using  those inputs is based on   the observation of DN$(k)$.
 Degrees of  most authors are around   their mode $5$, and only a few authors have a large degree. Hence the neighbors of an
author  are likely to be small degree authors. Therefore, for small degree authors, the degree differences between those authors and their neighbors are small,
and large for large degree authors,  which leads to
 the trend change  of DN$(k)$.

The correlation coefficient between hyperdegree     and the average hyperdegree of neighbors is around zero in each   science category~(Table~\ref{tab4}).
 For the authors with  $k$-hyperdegree,   denote the    average hyperdegree of their  neighbors        by    HN$(k)$. It means   choosing collaborators is free of the factor of    productivity.
In reality,
members of a research team may have various scientific ages (newcomers, incumbents),   so  different  hyperdegrees.
Since collaborations mainly happen in a research team,
collaborators of an author   could have various hyperdegrees, which appears as the stable trend  of  HN$(k)$.

 Based on the average value of HN$(k)$ larger than $2$, and 74.62\% authors only having one paper in the data, we can derive that a large fraction of authors collaborate with at least one author who has published a paper  in PNAS 1999--2013 to publish their first paper in the data. The proportions of these authors are 79.22\%, 71.17\% and 65.12\% in biological, physical and social sciences respectively. The proportions may be overestimated, because some of these  authors may publish papers in PNAS before 1999.

\subsection{Interdisciplinarity at discipline level}

The  co-category proportion     measures the activities of  interdisciplinary research.
   There are   49.2\%,   46.0\% and  7.3\% authors of social, physical and biological sciences  who   published  interdisciplinary papers.
 The common sense suggests that social scientists
 engage in  research  solitary. The    proportion of social sciences shows that the    common sense does not hold in PNAS.
   Referene\cite{Levitt3} also shows, there has been a move towards increased interdisciplinarity  in recent decades  in social sciences.

Above analysis process could be implemented to the second-class disciplines to obtain a high-resolution result. However some disciplines only have a few papers, e. g. only 17 papers of political science. So  the   analysis  for those disciplines  loses statistical meaning.
  Hence we took another perspective to analyze the interactions  among the second-class disciplines   by visualizing them as the  network in Fig.~\ref{fig22}.
The network  is  connected,   i.~e.
no discipline is isolated.
Top three  nodes of this network  in terms of   degrees and  those in terms of  betweenness centralities are Applied mathematics,  Chemistry  and Anthropology~(Table~\ref{table1}).
It means the
theories, methods and problems of those disciplines  are directly or indirectly used or studied by many    disciplines.
 For each first-class discipline, we
 contracted its second-class disciplines as one node, and calculated the betweenness centrality of the contracted node.
  Their betweenness centrality   (Biological sciences 47.51,
 Physical sciences	163.81, Social sciences	161.72)  support the above analysis.

\begin{figure}
\centering
\includegraphics[height=1.6 in,width=4.5     in,angle=0]{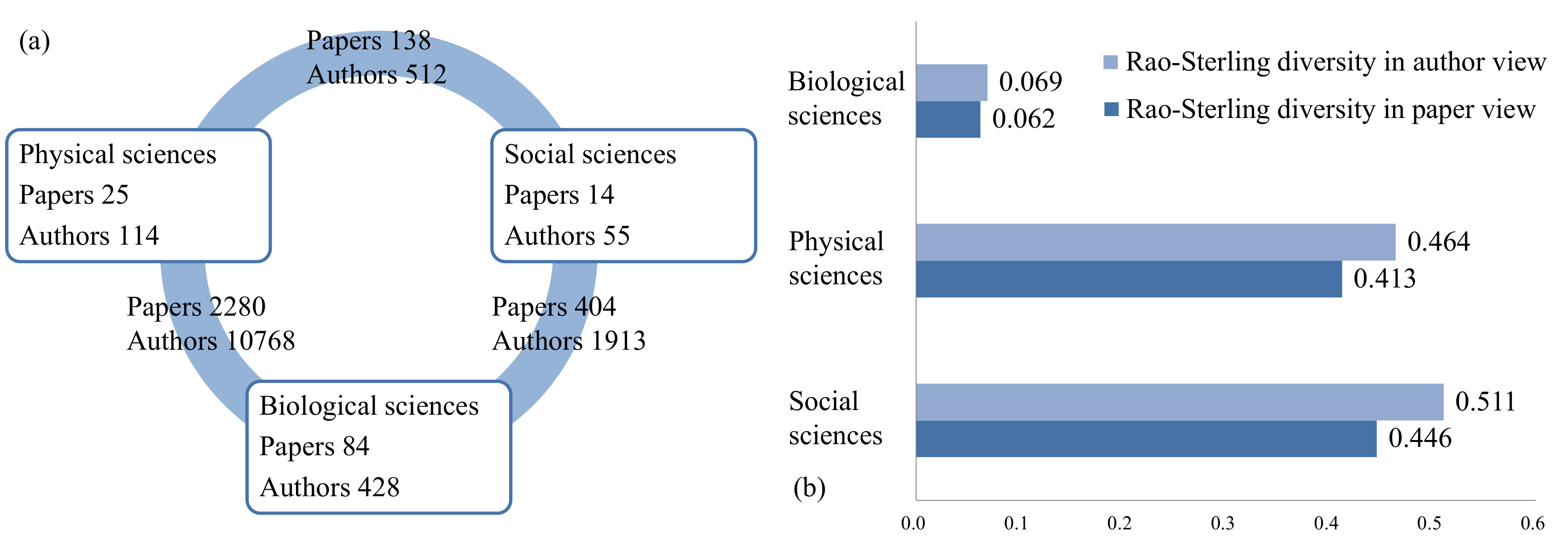}
 \caption{{\bf The interdisciplinary research of PNAS 1999--2013 between and within biological, physical and social sciences.} Panel (a) shows the
    proportions of papers and those of authors    involved in dyadic interactions
  between the three science categories, and those proportions
    involved in   interactions     within each science category.
 Panel (b) shows the Rao-Sterling diversity in paper/author view of  each science category, which
  measures the discipline diversity of interdisciplinary research.  }
 \label{fig6}      
\end{figure}

The co-category proportion only describes interdisciplinary activities.
Now let us measure the  discipline   diversity of interdisciplinary research in each science category through
   Rao-Sterling  index\cite{Stirling}
 $\Delta=\sum_{i,j(i\neq j)} d^\alpha_{ij} (p_ip_j)^\beta$, where $p_i$ and
 $p_j$ are proportional representations of the papers/authors in science
   category $i$ and $j$   and $d_{ij}$ is the level of difference attributed to  categories $i$ and $j$.
   Discipline information is used to classify authors into science categories: if one of his papers belongs to a discipline, an author can be classified into the discipline, so into the corresponding sciences. Note that an author can be classified into several science categories, if his papers belong to more than one discipline.
        Here  we let $\alpha=\beta=d_{ij}=1$ for all $i$ and $j$,
hence the calculated  Rao-Sterling index measures the balance-weighted variety of interdisciplinary research in the level of science categories.
  The   index in author view and that in paper view show that
 the   discipline diversity of interdisciplinary research in social sciences and that in physical sciences are much higher than that in
biological sciences (Fig.~\ref{fig6}).


\subsection{Interdisciplinarity at  author level}

\begin{figure}
\includegraphics[height=1.45   in,width=4.6    in,angle=0]{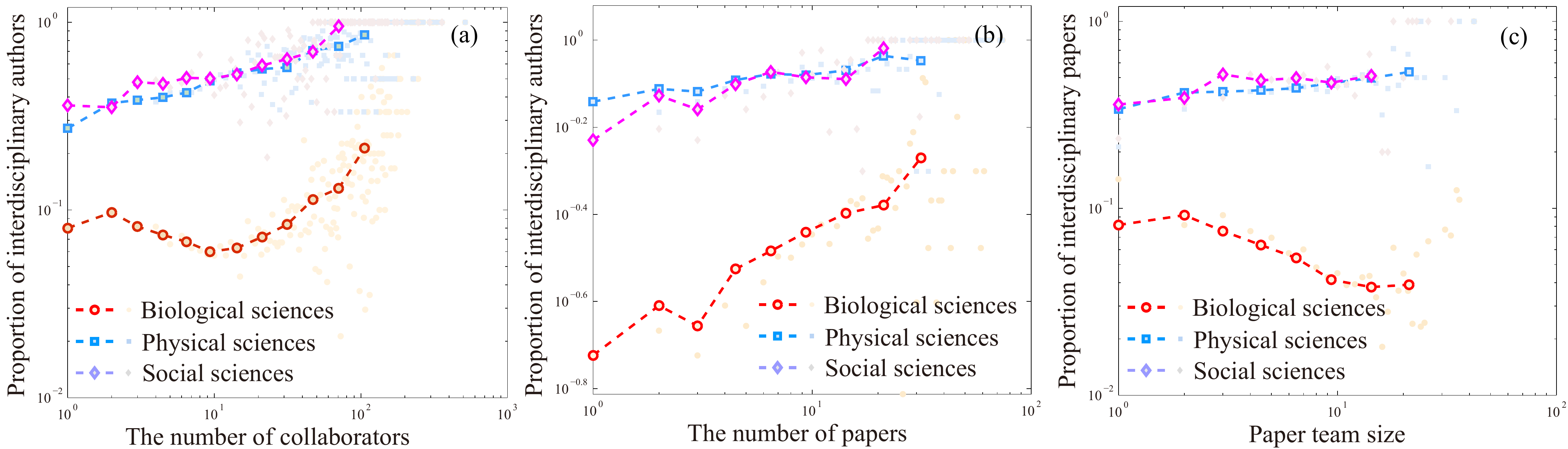}
 \caption{{\bf
 The relationship between authors' and papers'
specific indexes   and their interdisciplinarity.} Panels (a,b) show the relationship between author degree/hyperdegree and the
probability of doing interdisciplinary research.   Panel  (c)  shows the   relationship between
paper team size and the probability of being an interdisciplinary paper. }
 \label{fig8}      
\end{figure}

We analyzed  the relationship between   author degree/hyperdegree and the probability of doing interdisciplinary research,
 and the relationship
between paper team size and the    probability of being an interdisciplinary paper.
    Fig.~\ref{fig8}   shows  that in each science category,
 interdisciplinary research is not just carried out by authors with a large degree or those with a large hyperdegree.

             Fig.~\ref{fig8} also shows that  large degree or hyperdegree authors are  likely to   engage  in interdisciplinary research, and
 a paper  with a large  team size  is  likely to
  be  an interdisciplinary one.
 It seems  these phenomena   can  be expected at random.
  Take a set of elements (collaborators, papers) of several  classes, and select a subset randomly. Then   a larger subset more likely contains
  elements from more than one  class.
   This reasoning, though plausible, is incorrect, because
scientists do not randomly   select topic and collaborators. Research costs (investments of time and effort) make scientists  tend to   work  within  their familiar fields.
In addition, the  reasoning is based on that the selection scope of collaborators is  limited to   empirical  data, which does not hold in reality.

\begin{figure}
\includegraphics[height=3.6in,width=4.7    in,angle=0]{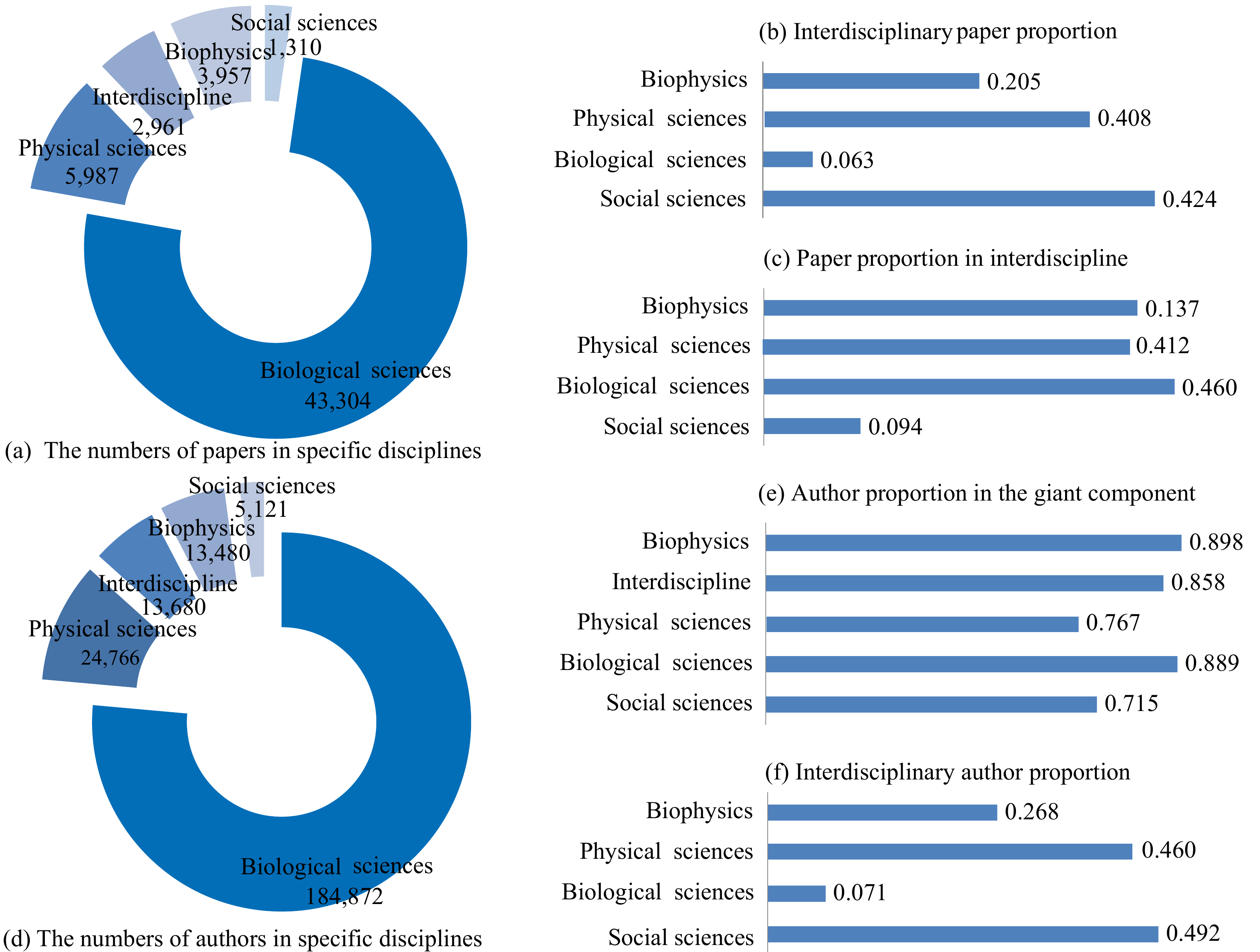}
 \caption{{\bf Interdisciplinary extents    of specific disciplines.}
 For each considered   discipline $i$, we denote its authors,  its authors involved in interdisciplinary research,
its   papers,  and its interdisciplinary papers
      by sets $A_i$, $A^I_i$, $P_i$ and $P^I_i$ respectively.  Denote   the giant component   of   coauthorship network  PNAS 1999-2013
 by $S$.
 The indexes are $|P_i|$ in Panel   (a),   $|P_i\cap I_i|/|I_i|$ in Panel   (b),  $|P_i\cap P^I_i|/|P^I_i|$ in Panel   (c), $|A_i|$  in Panel   (d),
 $|A_i\cap S|/|A_i|$ in   Panel (e), and $|A_i\cap A^I_i|/|A_i|$ in   Panel (f).
   }
 \label{fig5}      
\end{figure}

We analyzed the giant component of coauthorship network PNAS 1999-2013, which contains more than 86.8\% authors.
 There are 71.5\%,   76.7\% and  88.9\% authors of social, physical and biological sciences in the giant component~(Fig.~\ref{fig5}e).
  Note that the author misidentification caused
by initial-based methods increases the size of the ground-truth giant component\cite{Kim1}. Hence we identified authors by their provided names on papers (which
likely split one author into two) to obtain a conservative result.

\begin{figure}
\centering
\includegraphics[height=1.4in,width=4.     in,angle=0]{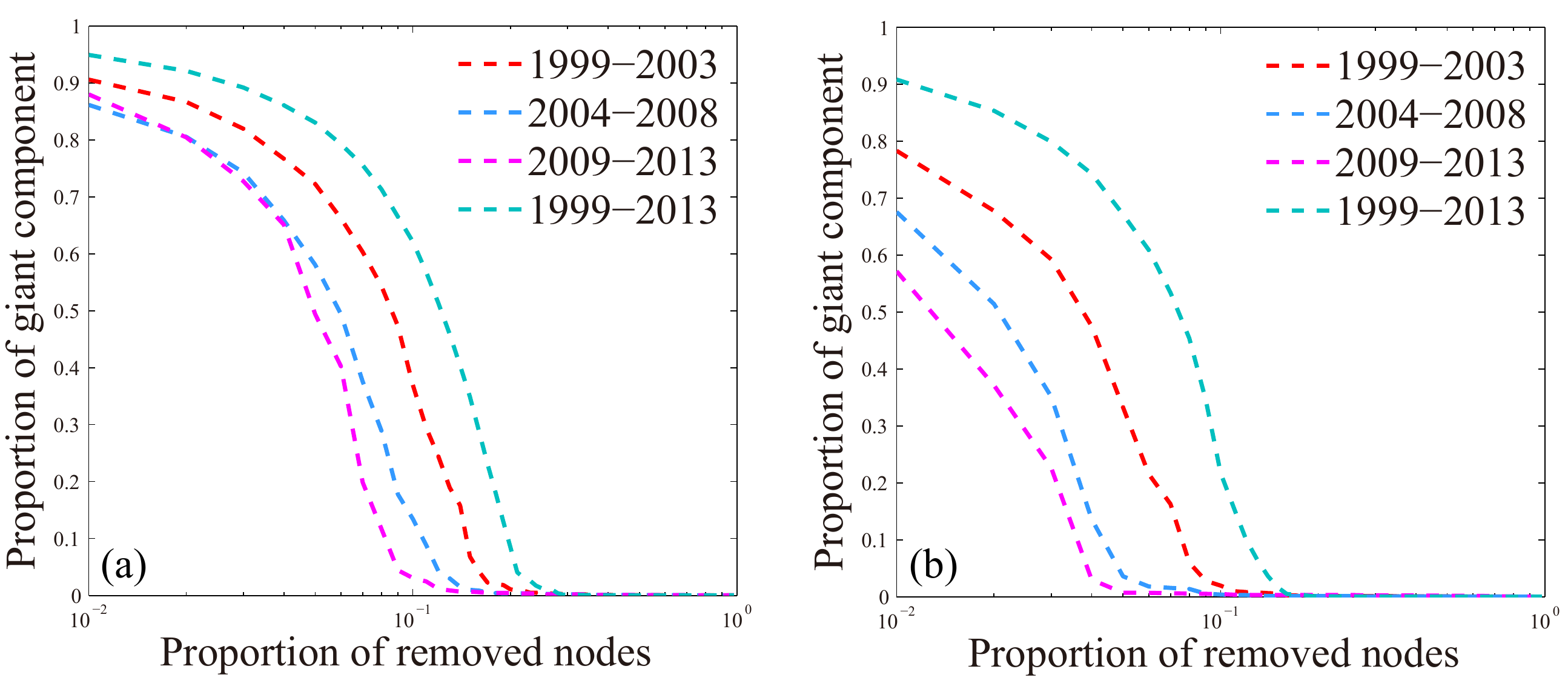}
 \caption{{\bf The relationship between   giant component size and degree/hyperdegree.} Nodes are removed from high degree/hyperdegree to low respectively. For degree and hyperdegree respectively, the relation curves between the proportion of removed nodes and that of the giant component show that a considerable number of authors contribute to the formation of giant component. The left-shifting trend of the relation curves in three time periods
 (1999--2003, 2004--2008 and 2009--2013) over time shows the increasing contributions of author activity and productivity to the formation of the giant component.
 }
 \label{fig7}      
\end{figure}

Interdisciplinary research and multidisciplinary research contribute to the giant component containing most authors of each science category. We analyzed the relationship between the author proportion of the giant component and author activity/productivity. Remove authors from high degree and hyperdegree to low respectively, and calculate the proportion of the giant component. From the relation curve between the proportion of removed authors and that of the giant component, we can find that the formation of giant component is contributed by a considerable number of authors, e. g. the top 10\% authors ranked by degree (Fig.~\ref{fig7}). Consider the relationship in three time periods, viz. 1999--2003, 2004--2008 and 2009--2013. The relation curve shifts to the left over time, which means author activity and productivity are playing increasingly important roles in the formation of the giant component.

\section{Discussion and conclusions}
Our  case study on  PNAS 1999-2013 verifies  the similar transitivity and assortativity of collaboration patterns  in biological, physical and social sciences. The data demonstrate that the degree distribution types  of   the three science categories are identical,    which are a mixture of a generalized Poisson distribution and a power-law.
This also holds for hyperdegree.
 We provided an    explanation  for the emergence of  this distribution type  through   authors' ``yes/no" decisions and
their different abilities to attract  collaborations.


The data show that a considerable number of authors pursue interdisciplinary research, and
  the
giant component  of   coauthorship network   PNAS 1999-2013  contains most  authors of each science category.
  We took  network perspective to analyze the interactions  among the second-class disciplines, and
quantify their interdisciplinarity by
  network indexes such as degree and betweenness centrality.  We  found that specific second-class disciplines  (such as Applied mathematics and Anthropology) play an important role in interdisciplinary research.

The case study
contributes to understanding   multidisciplinary and interdisciplinarity collaboration patterns, due to the importance   of PNAS   and   to the   accurate  discipline information of its papers.  The selection of data   might
affect the details of our   findings about interdisciplinarity. Our   results
 may not  be interpreted as   the patterns
of general  researchers.
 For example, we cannot expect to observe
a high extent of interdisciplinarity by analyzing a  domain specific journal.
We finished the case study by asking a question:
What are the grounds of   interdisciplinary research?
 While a thorough discussion of this question is
beyond the scope of this paper, the following provides a simple discussion.



  There is a tendency of   fragmentation for disciplines in the development of sciences: going to split into sub-disciplines and specific topics.
 Although the research objects   are different, their  research paradigms are in common, which can be  grouped into four categories, namely theoretical research, experiment, simulation, and data-driven\cite{Hey}.
Meanwhile,
many scientific problems are too complex to be   understood  through the methodology of single discipline.   Integrating   theoretical and methodological perspectives drawn
from   different  disciplines   creates a   unified methodology for research problems and  even vocabulary used to present concepts in specific disciplines\cite{Haythornthwaite}, which    drives the formation of  transdisciplinary disciplines\cite{Grauwin}.


Systems science, as a  typical transdisciplinary discipline,   studies   systems from simple to complex,    from natural to
 social sciences.
The parts of a system and      the   relations between parts   can be   abstracted as    networks.
The rapid development of  research on networks (model, algorithm,...) breeds  a new discipline, namely network science.
  Some researchers from biological, physical and social fields      investigate their respective problems under
network framework\cite{Brier}, e. g. our case study.

To follow up the above,  one would think that
 common  research  paradigms and methodology, especially those integrated as transdisciplinary disciplines,  give grounds for    the interactions between  science categories  and for the formation of    giant components in coauthorship networks.
   It seems promising that
  analyzing  paper content helps to
validate the universality of those paradigms and methodologies.
 Over half the    papers of PNAS 1999-2013 contain   the  topic words   ``system" and ``control"\cite{Xie4}.
  The high  proportion of
the papers containing a topic word at certain levels reflects the typicality of the   topic.
However, it is not easy to say which is the relation between a paper containing the word ``system" and a paper
    applying  research results of systems science.
Hence    validating the universality at semantic level is a subject for further study.

\section*{Availability of data and materials}

The   data are freely available from the their website  http://www.pnas.org.
 Feel free to get in contact with the corresponding
author in case you need more information.

\section*{Competing interests}
The authors declare that they have no competing interests.

\section*{Funding}
ZX acknowledges support from      National   Science Foundation of China (NSFC) Grant No. 61773020.

\section*{Authors' contributions}
All authors conceived and designed the research. ZX and ML wrote
the paper.  ZX  and JPL analyzed the data.   OYZZ  acquired the data.
ZX and XJD wrote the discussion.
 All authors
discussed the research  and approved the final version of the manuscript.

\section*{Acknowledgments}
We thank the   anonymous reviewers for their valuable suggestions and great help.

\section*{Endnotes}

$^a$http://www.pnas.org.

\noindent $^b$Wikipedia shows
 that people with major 100 Chinese surnames account  for   84.77\%  of   the total Chinese population.

%




%
%


\section{Appendix}


The following  boundary    detection  algorithms come from  Reference~\cite{Xie6}.
\begin{table*}[!ht] \centering \caption{{\bf A boundary detection algorithm for probability density functions.} }
\begin{tabular}{l r r r r r r r r r} \hline
Input: Observations  $D_s$ ($s=1,...,n$),  rescaling function $g(\cdot)$,  and fitting model     $h(\cdot)$.\\
\hline
For   $k$ from $1$ to $\max(D_1,...,D_n)$ do: \\
~~~~Fit   $h(\cdot)$  to   the PDF $h_0(\cdot)$ of    $\{D_s, s=1,...,n|D_s \leq k\}$    by  maximum-likelihood\\ estimation; \\
~~~~Do    KS test for two    data
     $g(h(t))$ and $g( h_0(t))$, $t=1,...,k$ \\
   with the null hypothesis  they coming from the same   distribution;\\
~~~~Break  if  the test rejects the null hypothesis  at     significance level $5\%$. \\ \hline
Output: The current $k$ as the   boundary point. \\ \hline
 \end{tabular}
    \begin{flushleft}
    \end{flushleft}
\label{tab5}
\end{table*}
\begin{table*}[!ht] \centering \caption{{\bf Boundary point  detection algorithm for general functions.} }
\begin{tabular}{l r r r r r r r r r} \hline
Input: Data vector  $h_0(s)$ ($s=1,...,K$), rescaling funtion $g(\cdot)$, and fitting model     $h(\cdot)$.\\
\hline
For   $k$ from $1$ to $K$ do: \\
~~~~Fit   $h(\cdot)$  to    $h_0(s)$, $s=1,...,k$  by regression; \\
~~~~Do  KS test for two    data vectors
     $g(h(s))$ and $g( h_0(s))$, $s=1,...,k$ with   the null\\
 hypothesis they coming from the same   distribution;\\
~~~~Break  if  the test rejects the null hypothesis  at   significance level  $5\%$. \\ \hline
Output: The current $k$ as the  boundary point. \\ \hline
 \end{tabular}
\label{tab6}
\end{table*}

\end{document}